\documentclass[usenatbib]{mn2e}

\usepackage{graphicx}
\usepackage{mathptmx}
\usepackage[hang,raggedright]{subfigure}
\graphicspath{{./Images/}}

\begin{document}
\title[Infrared-Faint Radio Sources in the xFLS]{Deep 610-MHz Giant
  Metrewave Radio Telescope observations of the {\it Spitzer}
  extragalactic First Look Survey field -- III. The radio properties
  of Infrared-Faint Radio Sources}

\author[T.\ Garn \& P.\ Alexander]{Timothy Garn\thanks{E-mail:
    tsg25@cam.ac.uk}, Paul Alexander
    \\Astrophysics Group, Cavendish Laboratory, 19
    J.~J.~Thomson Ave., Cambridge CB3~0HE}

\date{\today}
\pagerange{\pageref{firstpage}--\pageref{lastpage}; } \pubyear{2008}
\pubyear{2008}
\label{firstpage}
\volume{000}
\maketitle

\begin{abstract}
Infrared-Faint Radio Sources (IFRSs) are a class of source which are
bright at radio frequencies, but do not appear in deep infrared
images.  We report the detection of 14 IFRSs within the {\it Spitzer}
extragalactic First Look Survey field, eight of which are detected
near to the limiting magnitude of a deep $R$-band image of the region,
at $R\sim24.5$.  Sensitive {\it Spitzer Space Telescope} images are
stacked in order to place upper limits on their mid-infrared flux
densities, and using recent 610-MHz and 1.4-GHz observations we find
that they have spectral indices which vary between $\alpha=0.05$ and
1.38, where we define $\alpha$ such that $S_{\nu} =
S_{0}\nu^{-\alpha}$, and should not be thought of as a single source
population.  We place constraints on the luminosity and linear size of
these sources, and through comparison with well-studied local objects
in the 3CRR catalogue demonstrate that they can be modelled as being
compact ($<20$~kpc) Fanaroff-Riley Class II (FRII) radio galaxies
located at high redshift ($z\sim4$).
\end{abstract}


\section{Introduction}
Infrared-Faint Radio Sources (IFRSs) are a class of radio-bright or
infrared-faint objects identified in the Australia Telescope Large
Area Survey \citep*[ATLAS;][]{Norris06}.  The ATLAS survey consists of
deep 1.4-GHz radio observations over two of the southern {\it Spitzer}
Wide-area Infrared Extragalactic Survey \citep[SWIRE;][]{Lonsdale03}
fields, which have deep infrared data between 3.6 and 160~$\mu$m
publicly available.  The {\it Spitzer Space Telescope}
\citep{Werner04} infrared observations of the SWIRE fields are
sensitive enough that it was expected that all radio sources detected
in the ATLAS survey which were in the local Universe would be visible
in the SWIRE images \citep{Norris06}, whether the radio emission was
due to star-formation or an Active Galactic Nuclei (AGN).  However, it
was found that some radio sources did not have an infrared counterpart
at any of the {\it Spitzer} wavelengths.  A total of 55 IFRSs of
varying flux densities have been detected within the {\it Chandra}
Deep Field South \citep[CDFS;][]{Norris06} and European Large-Area
{\it ISO} Survey South-1 \citep[ELAIS-S1;][]{Middelberg08} fields,
with several having 1.4-GHz flux densities of greater than 5~mJy, and
the brightest having a 1.4-GHz flux density of 26.1~mJy
\citep{Norris06}.

The possibility that these sources are obscured star-forming galaxies
with infrared flux densities falling just below the sensitivity limit
of {\it Spitzer} would seem to be ruled out by the production of
stacked infrared images centred on the radio source positions of 22
IFRSs \citep{Norris06} which show no evidence for any infrared
counterparts.  \citet{Norris07} made observations of two of their
bright IFRSs using Very Long Baseline Interferometry (VLBI), and found
that one of them had a compact core with angular size less than
0.03~arcsec (a linear size below 260~pc at any reasonable redshift),
although they were unable to detect the second target.  They concluded
that the radio emission is driven by AGN activity rather than
star-formation, making IFRSs either high redshift radio-loud quasars,
or abnormally highly obscured radio galaxies at more moderate
redshifts.  \citet{Norris07} considered the spectral energy
distribution of the radio-loud quasar 3C273, and showed that placing
it at low redshift could not reproduce both the detection at 1.4~GHz
and non-detection at 3.6~$\mu$m for the source with the compact core,
but moving 3C273 to $z=7$ would fulfil the requirements.  However, the
lack of information at other wavelengths makes it difficult to
determine exactly what type of source the IFRSs are. These sources are
selected through being radio-bright, making the use of radio
diagnostics important for determining their type.

The first of the large-scale surveys carried out by the {\it Spitzer
Space Telescope} was the {\it Spitzer} extragalactic First Look Survey
(xFLS).  Four square degrees, centred on $17^{\rm h}18^{\rm m}00^{\rm
s}$, $+59\degr30'00''$ (J2000 coordinates, which are used throughout
this work), were observed with two instruments -- the Infrared Array
Camera \citep[IRAC;][]{Fazio04} at 3.6, 4.5, 5.8 and 8~$\mu$m, and the
Multiband Imaging Photometer for {\it Spitzer}
\citep[MIPS;][]{Rieke04} at 24, 70 and 160~$\mu$m.  This data is
comparable in depth to the SWIRE survey, with catalogue completeness
limits ($5\sigma$, apart from the 3.6-$\mu$m band which has its
completeness limit set at $7\sigma$) of 20, 25, 100 and 100~$\mu$Jy in
the four IRAC bands.  There are two deep radio surveys of the region,
at 1.4~GHz with the Very Large Array \citep[VLA;][]{Condon03} and at
610~MHz with the Giant Metrewave Radio Telescope
\citep[GMRT;][hereafter Paper I]{Garn07}.  These have comparable
resolution (5~arcsec$^{2}$ at 1.4~GHz; $5.8\times4.7$~arcsec$^{2}$ at
610~MHz) and sensitivity (23~$\mu$Jy~beam$^{-1}$ at 1.4~GHz;
30~$\mu$Jy~beam$^{-1}$ at 610~MHz), and cover approximately the same
region of sky as the infrared observations.  An $R$-band survey of the
xFLS field has been carried out \citep{Fadda04} with a limiting Vega
magnitude of $R$ = 25.5, and estimated 50~per~cent completeness limit
of $R$ = 24.5.  The deep and complementary infrared, optical and radio
observations make this region a good area to search for other IFRSs,
and the availability of multi-frequency radio data will allow the
spectral index $\alpha$\footnote{where we define the radio spectral
index $\alpha$ such that $S_{\nu} = S_{0} \nu^{-\alpha}$} of any IFRS
to be obtained, providing a much greater understanding of the source
properties.

In section~\ref{sec:selection} we describe the selection criteria for
forming a sample of Infrared-Faint Radio Sources.  We consider the
available radio and infrared properties of these sources in
section~\ref{sec:properties}, create stacked infrared images of the
sources in order to test whether the sources lie slightly below the
{\it Spitzer} detection threshold, and place upper limits on their
infrared flux between 3.6 and 160~$\mu$m.  We calculate the radio
spectral index of each source, and demonstrate that IFRSs are made up
of flat, steep and ultra-steep spectrum sources.  In
section~\ref{sec:modelling} we place limits on the luminosity and
linear size of these sources, and compare well-studied sources from
the Revised Revised Third Cambridge (3CRR) catalogue of
\citet*{Laing83} to the observed IFRS population.  We demonstrate that
it is possible to model the flat, steep and ultra-steep-spectrum IFRSs
with the spectral energy distribution (SED) of a local radio galaxy,
placed at high redshift, and conclude that it is possible to explain
the IFRS population without requiring new, exotic source types.  A
flat cosmology with the best-fitting parameters from the Wilkinson
Microwave Anisotropy Probe (WMAP) five-year data of $\Omega_{\Lambda}
= 0.74$ and H$_{0}$ = 72~km~s$^{-1}$~Mpc$^{-1}$ \citep{Dunkley08} is
assumed throughout this work.

\section{Sample selection}
\label{sec:selection}
\subsection{Initial radio source sample}
The xFLS region was observed at 1.4~GHz by \citet{Condon03} with the
Very Large Array (VLA), covering $\sim4$~deg$^{2}$ with a noise level
of 23~$\mu$Jy~beam$^{-1}$.  We created a sample of radio sources at
1.4~GHz within the region covered by the VLA and GMRT surveys, using
{\sc Source Extractor} \citep[{\sc SExtractor};][]{Bertin96} and the
technique described in Paper I.  A peak flux density of greater than
0.5~mJy was required for all sources, equivalent to a signal-to-noise
ratio of $\approx20$.  This ensures that the sources are
unquestionably real, and that sources which are found to be
infrared-faint will be extreme examples of their class -- the
inclusion of fainter radio sources in the sample would reduce the
significance of the lack of an infrared detection.  Within the region
covered by the VLA and GMRT surveys, 511 radio sources were detected.
We reject 107 of these for being located outside the coverage area of
one or more of the infrared images, leaving 404 1.4-GHz radio sources.

A similar catalogue of radio sources was created from the 610-MHz
image, requiring sources to be detected at the $5\sigma$ level but
with no further selection criteria.  At the edges of the 610-MHz
image, $5\sigma$ corresponds to a flux density of
$\sim750$~$\mu$Jy~beam$^{-1}$, decreasing to
$\sim150$~$\mu$Jy~beam$^{-1}$ towards the pointing centres.  Sources
at the 1.4-GHz peak flux limit will therefore be seen anywhere in the
610-MHz survey if they have a spectral index $\alpha$ greater than
0.5, so some flat-spectrum 1.4-GHz sources may not be assigned 610-MHz
counterparts using this criterion.

The majority of the 1.4-GHz sources (362/404; 90~per~cent) had a
610-MHz counterpart within 1.5~arcsec, equivalent to one pixel in the
radio images.  A further 24 had counterparts within 3~arcsec.  The
remaining 18 sources were inspected in order to identify why no
610-MHz counterpart was found.  Seven were either one component of a
double radio source, or extended radio sources with 610-MHz
counterparts $>3$~arcsec away; two were close to a bright source and
were concealed by the increased noise surrounding it (see Paper~I for
details on the effects of bright sources on the 610-MHz image); the
remaining nine were near to the edge of the 610-MHz mosaic, and not
detected due to the increased noise due to the decreasing gain of the
GMRT primary beam.  All 18 sources had an infrared counterpart (see
section~\ref{sec:removeIR}), and were rejected from the sample,
leaving 386 sources to make up the initial `radio-bright' sample.

\subsection{Removing infrared counterparts}
\label{sec:removeIR}
\begin{figure}
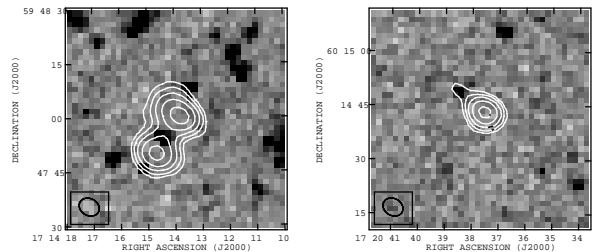

  \centerline{ \subfigure[A double radio source, with infrared
    counterpart lying between the two radio lobes.]{
    \includegraphics[width=0.22\textwidth]{PLVER25.PS}}
    \subfigure[A radio source with slightly extended emission covering
    an infrared counterpart. ]{
    \includegraphics[width=0.22\textwidth]{PLVER71.PS}}
    }
\caption{Two examples of radio sources with catalogued infrared
  counterparts which were not initially classified as being a match.
  Contours are from the 610-MHz data, and are 0.5, 1, 2, 4,
  8~mJy~beam$^{-1}$, while the grey-scale shows the 3.6-$\mu$m image,
  and ranges between $\pm$ the 3.6-$\mu$m noise level of
  2.9~$\mu$Jy~pixel$^{-1}$.  }
\label{fig:source1}
\end{figure}

We follow \citet{Norris06} in rejecting sources from the radio-bright
sample if they have an infrared counterpart within 3~arcsec in any of
the published {\it Spitzer} source catalogues
\citep{Lacy05,Fadda06,Frayer06}.  This generated an initial list of 94
IFRS candidates, which required visual inspection in order to check
for the presence of infrared counterparts which had not been
identified as being related to the radio sources.  Composite
infrared-radio maps were generated for each source, after rotating the
geometry of the IRAC images to match the radio observations.  The
majority of the candidate sources (49/94; 52~per~cent) were definite
components of double or triple radio galaxies with an infrared
counterpart present, but not co-located with the radio emission --
Fig.~\ref{fig:source1}a shows an example of one of these sources, with
610-MHz contours overlaid on the 3.6-$\mu$m image.  The counterpart is
sufficiently far from the centres of the radio components that it was
not classified as a match, but is clearly associated with the double
radio galaxy.  Four sources were located close to very bright infrared
sources, which would have concealed any true infrared counterparts.
There were 21 radio sources which should have been matched to a source
in the IRAC catalogue, due to either having extended radio emission
which covered an infrared source, or having a slightly greater
separation between the infrared and radio source centres than 3~arcsec
due to the lower resolution of the radio images --
Fig.~\ref{fig:source1}b shows an example of one of these sources.  One
radio source showed no evidence for infrared emission, but was in a
noisy region of the 610-MHz image, and the measurement of the radio
flux was very uncertain.  After rejecting all of these types of
source, there were 19 remaining radio sources with no counterparts
within the infrared catalogues.

\begin{figure}
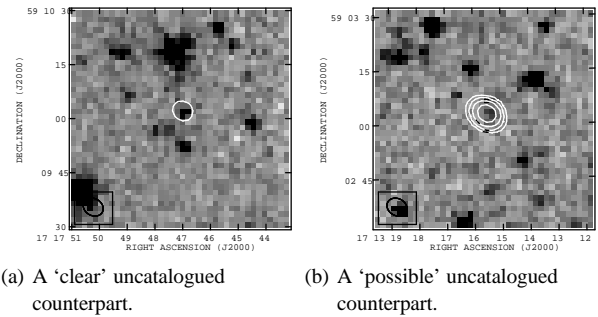

  \centerline{
    \subfigure[A `clear' uncatalogued counterpart.]{
      \includegraphics[width=0.22\textwidth]{PLVER51.PS}}
    \subfigure[A `possible' uncatalogued counterpart.]{
      \includegraphics[width=0.22\textwidth]{PLVER18.PS}}
  }
\caption{Two radio sources with uncatalogued infrared
  counterparts. Contours are from the 610-MHz data, and are 0.5, 1, 2,
  4, 8~mJy~beam$^{-1}$, while the grey-scale shows the 3.6-$\mu$m
  image, and ranges between $\pm$ the 3.6-$\mu$m noise level of
  2.9~$\mu$Jy~pixel$^{-1}$.  }
\label{fig:counterparts}
\end{figure}

\begin{figure*}
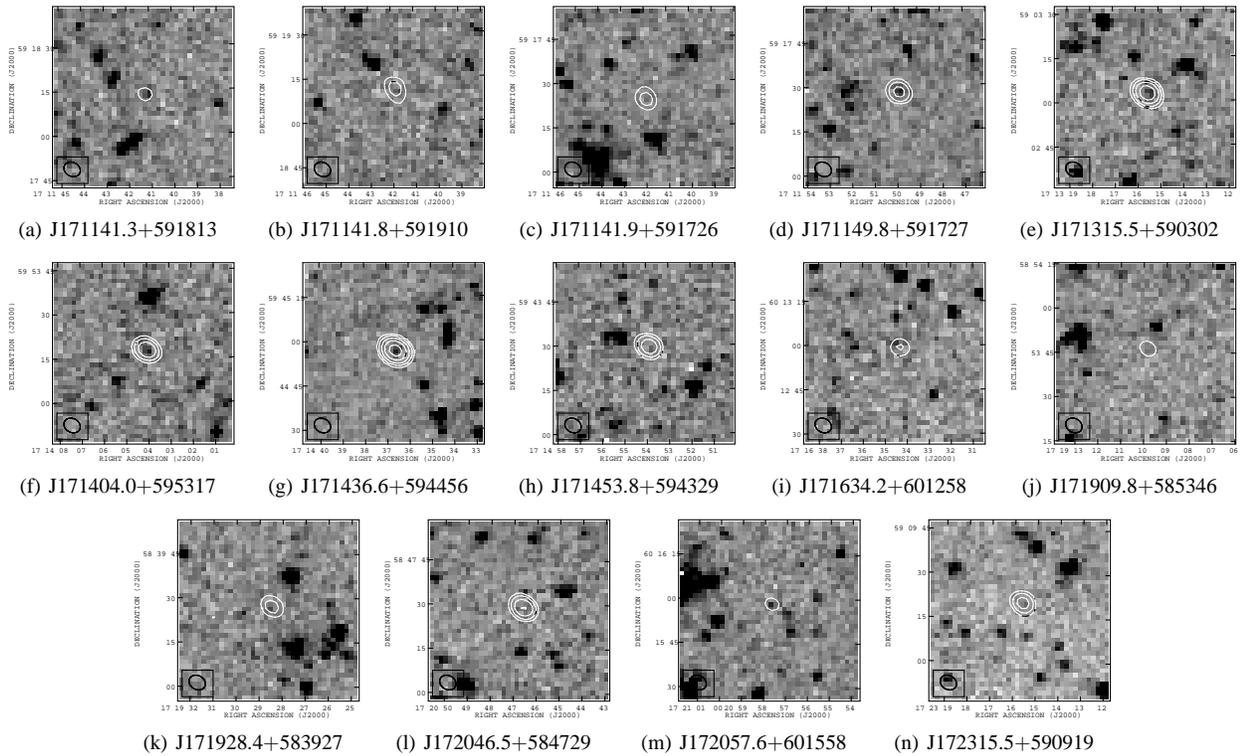

  \centerline{
    \subfigure[J171141.3$+$591813]{
      \includegraphics[width=0.18\textwidth]{PLVER6.PS}}
    \subfigure[J171141.8$+$591910]{
      \includegraphics[width=0.18\textwidth]{PLVER7.PS}}
    \subfigure[J171141.9$+$591726]{
      \includegraphics[width=0.18\textwidth]{PLVER8.PS}}
    \subfigure[J171149.8$+$591727]{
      \includegraphics[width=0.18\textwidth]{PLVER10.PS}}
    \subfigure[J171315.5$+$590302]{
      \includegraphics[width=0.18\textwidth]{PLVER18.PS}}
}
  \centerline{
    \subfigure[J171404.0$+$595317]{
      \includegraphics[width=0.18\textwidth]{PLVER24.PS}}
    \subfigure[J171436.6$+$594456]{
      \includegraphics[width=0.18\textwidth]{PLVER30.PS}}
    \subfigure[J171453.8$+$594329]{
      \includegraphics[width=0.18\textwidth]{PLVER34.PS}}
    \subfigure[J171634.2$+$601258]{
      \includegraphics[width=0.18\textwidth]{PLVER47.PS}}
    \subfigure[J171909.8$+$585346]{
      \includegraphics[width=0.18\textwidth]{PLVER58.PS}}
}
  \centerline{
    \subfigure[J171928.4$+$583927]{
      \includegraphics[width=0.18\textwidth]{PLVER62.PS}}
    \subfigure[J172046.5$+$584729]{
      \includegraphics[width=0.18\textwidth]{PLVER72.PS}}
    \subfigure[J172057.6$+$601558]{
      \includegraphics[width=0.18\textwidth]{PLVER74.PS}}
    \subfigure[J172315.5$+$590919]{
      \includegraphics[width=0.18\textwidth]{PLVER88.PS}}
}
  \caption{3.6-$\mu$m grey-scale images of the 14 IFRSs, with 610-MHz
  radio contours overlaid.  The designation of each source from the
  610-MHz FLSGMRT catalogue is given below each subfigure.  The
  grey-scale ranges between $\pm\sigma$, where $\sigma$ is the noise
  level of 2.9~$\mu$Jy~pixel$^{-1}$, in order to accentuate faint
  infrared emission.  The radio contours are at 0.5, 1, 2, 4 and
  8~mJy~beam$^{-1}$.  The resolution of the radio image is shown by
  the ellipse in the lower left of each image.}
  \label{fig:IFRSs}
\end{figure*}

Five of these sources showed clear evidence for an uncatalogued
infrared counterpart in the 3.6-$\mu$m image, with four of these also
showing evidence for an uncatalogued counterpart in the 4.5-$\mu$m
image.  Fig.~\ref{fig:counterparts}a shows an example of one of the
radio sources with a clear 3.6-$\mu$m counterpart, which was not
included in the IRAC catalogue.  We reject these five sources from the
sample.  A further six sources show possible evidence for an
uncatalogued infrared counterpart, with a slight increase in infrared
emission being seen at or close to the centre of the radio emission.
We retain these sources in the sample, but identify them as having
possible counterparts in the remainder of this analysis -- we will
test for a statistical increase in infrared flux at the centre of each
radio source in section~\ref{sec:ifrsstacking}.  There are 14 IFRSs
remaining in the sample, which are shown in Fig.~\ref{fig:IFRSs}.  All
images are $60\times60$~arcsec$^{2}$ in order to look for extended
structure.  The grey-scale ranges between $\pm\sigma$ for the
3.6-$\mu$m image noise level of $\sigma=2.9$~$\mu$Jy~pixel$^{-1}$ in
order to be sensitive to faint infrared emission.  The 1.4-GHz
contours, and 4.5, 5.8 and 8-$\mu$m grey-scale have a similar
morphology.  The properties of the IFRSs are given in
Table~\ref{tab:IFRSproperties}, and throughout this work, designations
of the IFRSs will be taken from the 610-MHz FLSGMRT catalogue from
Paper~I.

\subsection{Comparison with previous samples}
Two previous samples of IFRSs have been constructed.  \citet{Norris06}
found 22 IFRSs in the Australia Telescope (AT) observations of the
3.7~deg$^{2}$ CDFS field, while \citet{Middelberg08} found 31 IFRSs in
the observations of the 3.9~deg$^{2}$ ELAIS-S1 field.  The source
density of bright IFRSs (with 1.4-GHz flux greater than 1~mJy) is
comparable in both studies to the sample presented in this work, which
has a density of $\sim2.5$~deg$^{-2}$, although since we have set a
lower limit to the radio flux for our sample selection, we find fewer
faint sources.  While large-area and relatively shallow radio surveys
such as the 1.4-GHz Faint Images of the Radio Sky at Twenty-cm
\citep*[FIRST;][]{Becker95} survey could be used to find bright IFRSs,
the requirement for deep infrared observations means that future
studies of this type of source are still likely to be limited to
smaller well-studied regions such as the xFLS and SWIRE fields.

The number of radio sources which are detected increases rapidly with
decreasing flux density \citep*[e.g.][]{Seymour04}, with the radio sky
being dominated by AGN sources above $\sim1$~mJy at 1.4~GHz, and the
contribution from star-forming galaxies becoming important below
$\sim1$~mJy .  The fact that we do not see a rapid increase in the
number of IFRSs at lower radio fluxes is a further indication that
these sources are unlikely to be obscured star-forming galaxies.

The resolution of the AT radio images, at $11\times5$~arcsec$^{2}$
\citep{Norris06} or $10\times7$~arcsec$^{2}$ \citep{Middelberg08}, is
poorer than either the VLA or GMRT observations used in this work
(5~arcsec$^{2}$ and $5.8\times4.7$~arcsec$^{2}$ respectively), while
the SWIRE infrared observations used in the CDFS and ELAIS-S1 studies
are slightly more sensitive than the xFLS observations.  While both of
these effects may slightly alter the selection biases behind the
different IFRS samples, we have followed a similar selection procedure
to that described in \citet{Norris06} and believe that the sources
identified in the three works should be comparable.

\section{Infrared-Faint Radio Source properties}
\label{sec:properties}
\subsection{Source stacking}
\label{sec:ifrsstacking}
A `stacked' infrared image will reveal whether the IFRS population is
made up of sources that have infrared fluxes which are slightly below
the detection threshold.  If this is the case, the stacked images will
show a faint infrared source co-located with the radio source centres.
A stacking experiment implicitly assumes that the sources being
stacked are all of the same type -- in section~\ref{sec:spectralindex}
we demonstrate that this is not the case, with the IFRS population
being made up of sources with flat, steep and ultra-steep radio
spectra.  However, stacked images will still give an indication of
whether there is some infrared flux present which is just below the
detection limit for individual sources.

\citet{Norris06} looked at the mean flux in a series of stacked IRAC
images, and found no detections of IFRSs.  However, it has been shown
\citep[e.g.][]{White07,Beswick08,Garn08stacking} that a median stacked
image is more reliable than a mean image, since the median image is
more robust to the presence of a few outlier pixels.  We created 14
`cut-out' infrared images with size $60\times60$~arcsec$^{2}$, centred
on the radio source positions of the IFRS shown in
Fig.~\ref{fig:IFRSs}.  The infrared background varies slightly across
the field \citep{Lacy05} and was calculated using {\sc SExtractor} and
subtracted from each cut-out image.  Each pixel of the stacked image
was then created by calculating the median value of that pixel from
the 14 cut-out images.

\begin{table*}
\begin{center}
\caption{The Infrared-Faint Radio Sources within the {\it Spitzer}
    extragalactic First Look Survey field.  Column~1 gives the
    designation of each source from the 610-MHz catalogue of
    \citet{Garn07}, columns~2 and 3 give the 1.4-GHz and 610-MHz flux
    densities, column~4 gives the radio spectral index $\alpha$,
    column~5 gives the $R$-band magnitude of each source from
    \citet{Fadda04}, column~6 gives the deconvolved angular diameters
    of each source from \citet{Condon03} and column~7 identifies those
    sources which have been classified as having potential
    uncatalogued infrared counterparts.}
\label{tab:IFRSproperties}
\begin{tabular}{c|c|c|c|c|c|c}
\hline
FLSGMRT & 1.4-GHz flux & 610-MHz flux & $\alpha$ & $R$-band &
$\theta$ & Possible\\
designation & (mJy)        & (mJy)        & & (Vega mag.) & (arcsec) &
counterpart\\
(1) & (2) & (3) & (4) & (5) & (6) & (7) \\
\hline 
J171141.3$+$591813 & $0.571\pm0.041$ & $0.907\pm0.075$ & $0.56\pm0.13$ & $23.26\pm0.12$
& $<3.1\times<2.3$ & Y\\ 
J171141.8$+$591910 & $1.074\pm0.058$ & $2.376\pm0.107$ & $0.96\pm0.08$ & $23.52\pm0.10$
& $~~~~4.9\times~~~~3.6$ & \\
J171141.9$+$591726 & $1.059\pm0.061$ & $2.298\pm0.113$ & $0.93\pm0.09$ & $24.17\pm0.14$
& $~~~~4.0\times<2.6$ & \\
J171149.8$+$591727 & $1.400\pm0.043$ & $3.181\pm0.093$ & $0.99\pm0.05$ & $>24.5$
& $<1.8\times<1.8$ & Y\\
J171315.5$+$590302 & $2.578\pm0.055$ & $8.024\pm0.134$ & $1.37\pm0.03$ & $24.31\pm0.24$
& $<1.8\times<1.8$ & Y\\
J171404.0$+$595317 & $2.484\pm0.042$ & $4.403\pm0.080$ & $0.69\pm0.03$ & $>24.5$
& $<1.9\times<1.9$ & Y\\
J171436.6$+$594456 & $4.108\pm0.044$ & $12.87\pm0.118$ & $1.38\pm0.02$ & $23.31\pm0.11$
& $<1.9\times<1.8$ & Y\\
J171453.8$+$594329 & $3.738\pm0.043$ & $3.886\pm0.087$ & $0.05\pm0.03$ & $24.70\pm0.15$
& $<1.9\times<1.8$ & \\
J171634.2$+$601258 & $1.341\pm0.045$ & $1.472\pm0.084$ & $0.11\pm0.07$ & $24.61\pm0.23$
& $<1.8\times<1.8$ & \\
J171909.8$+$585346 & $0.834\pm0.052$ & $1.057\pm0.079$ & $0.29\pm0.12$ & $>24.5$
& $<2.0\times<2.0$ & \\
J171928.4$+$583927 & $0.813\pm0.060$ & $1.928\pm0.023$ & $1.04\pm0.09$ & $24.21\pm0.20$
& $<2.0\times<2.0$ & \\
J172046.5$+$584729 & $3.705\pm0.065$ & $4.967\pm0.018$ & $0.35\pm0.05$ & $>24.5$
& $<1.8\times<1.7$ & \\
J172057.6$+$601558 & $0.654\pm0.039$ & $0.779\pm0.099$ & $0.21\pm0.17$ & $>24.5$
& $<2.9\times<2.2$ & Y\\
J172315.5$+$590919 & $1.531\pm0.049$ & $3.194\pm0.092$ & $0.88\pm0.05$ & $>24.5$
& $<2.3\times<2.2$ & \\
\hline
\end{tabular}
\end{center}
\end{table*}

\begin{figure*}
  \centerline{\subfigure[3.6-$\mu$m image, $N=14$.]{
\includegraphics[width=0.22\textwidth]{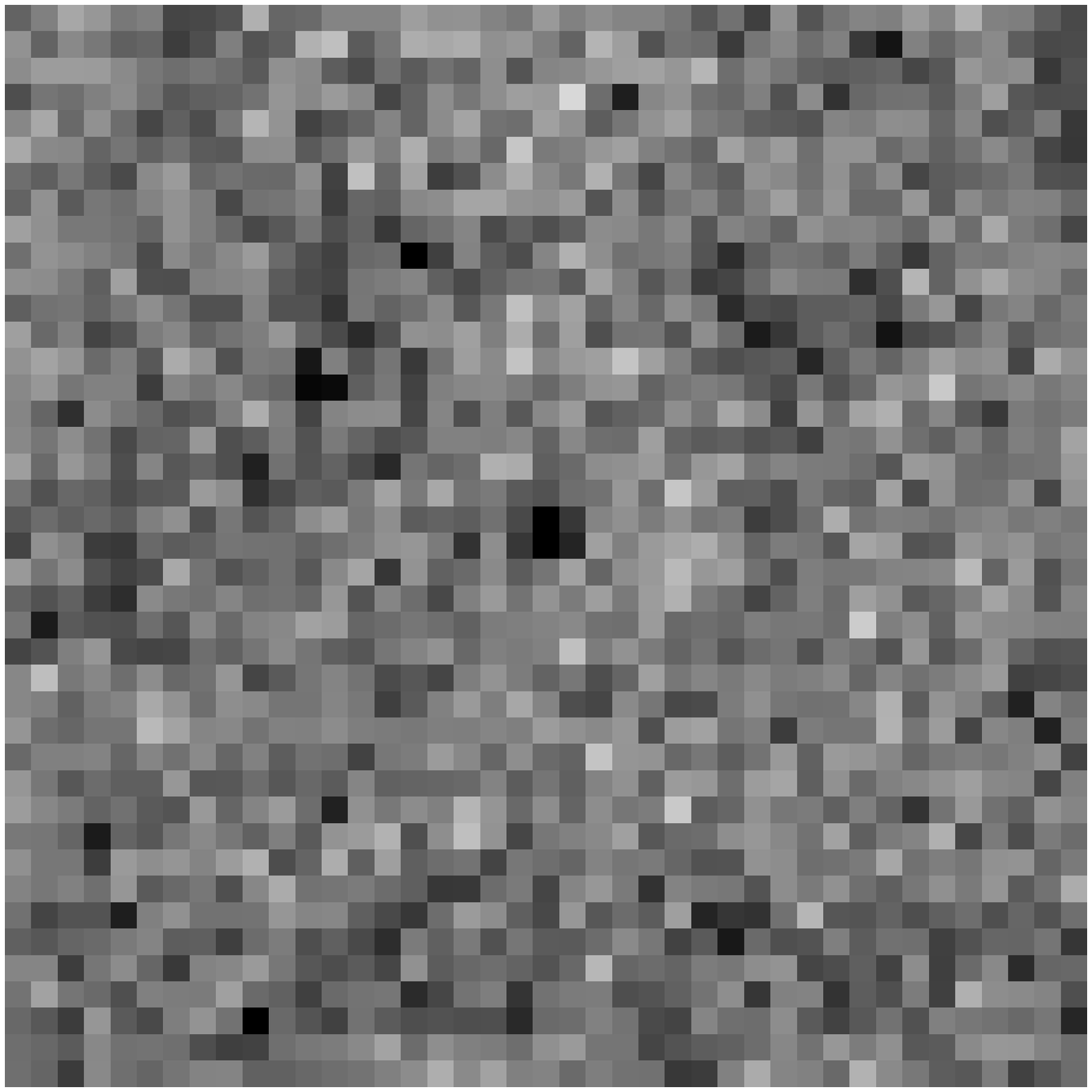}}
              \subfigure[4.5-$\mu$m image, $N=14$.]{
\includegraphics[width=0.22\textwidth]{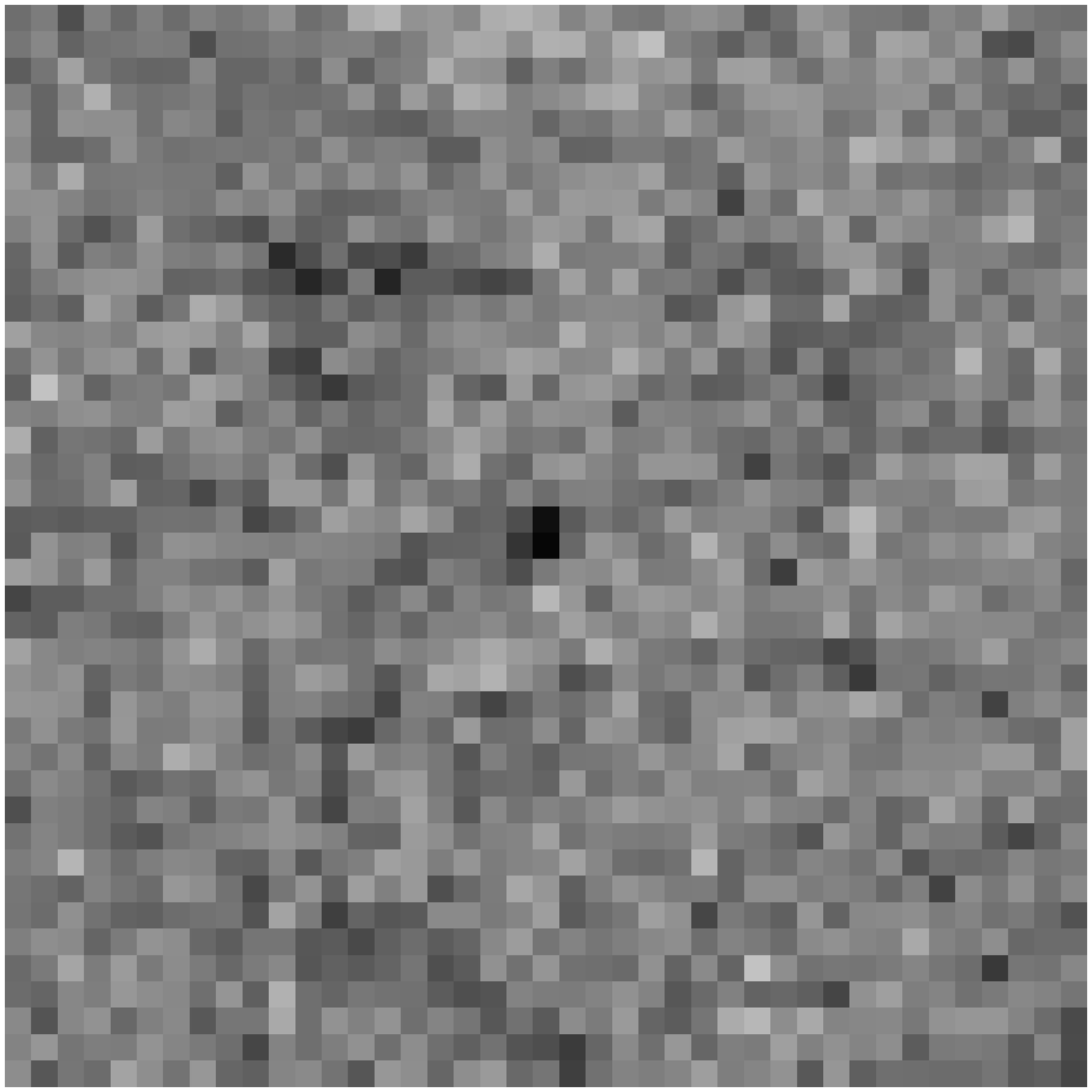}}
              \subfigure[5.8-$\mu$m image, $N=14$.]{
\includegraphics[width=0.22\textwidth]{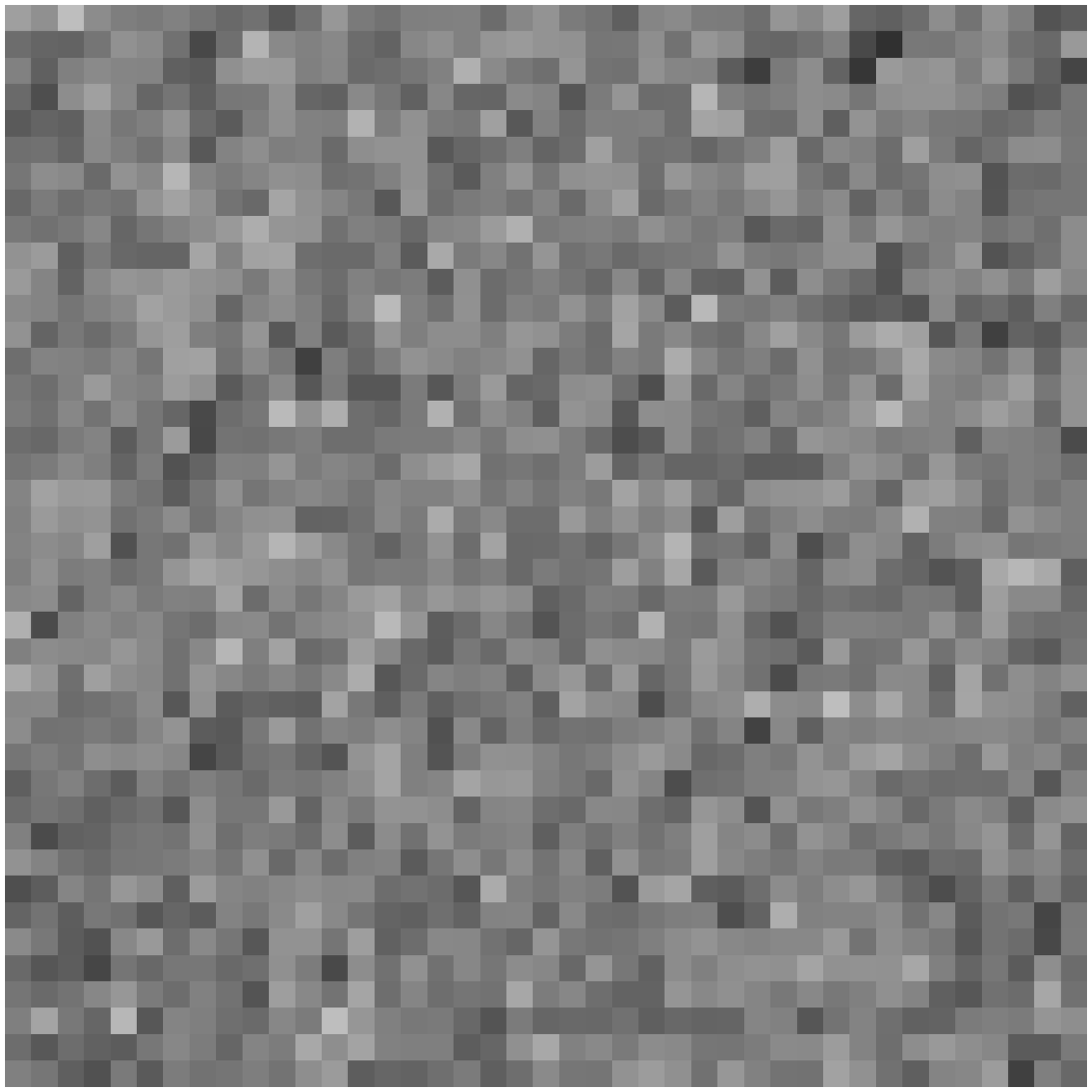}}
              \subfigure[8-$\mu$m image, $N=14$.]{
\includegraphics[width=0.22\textwidth]{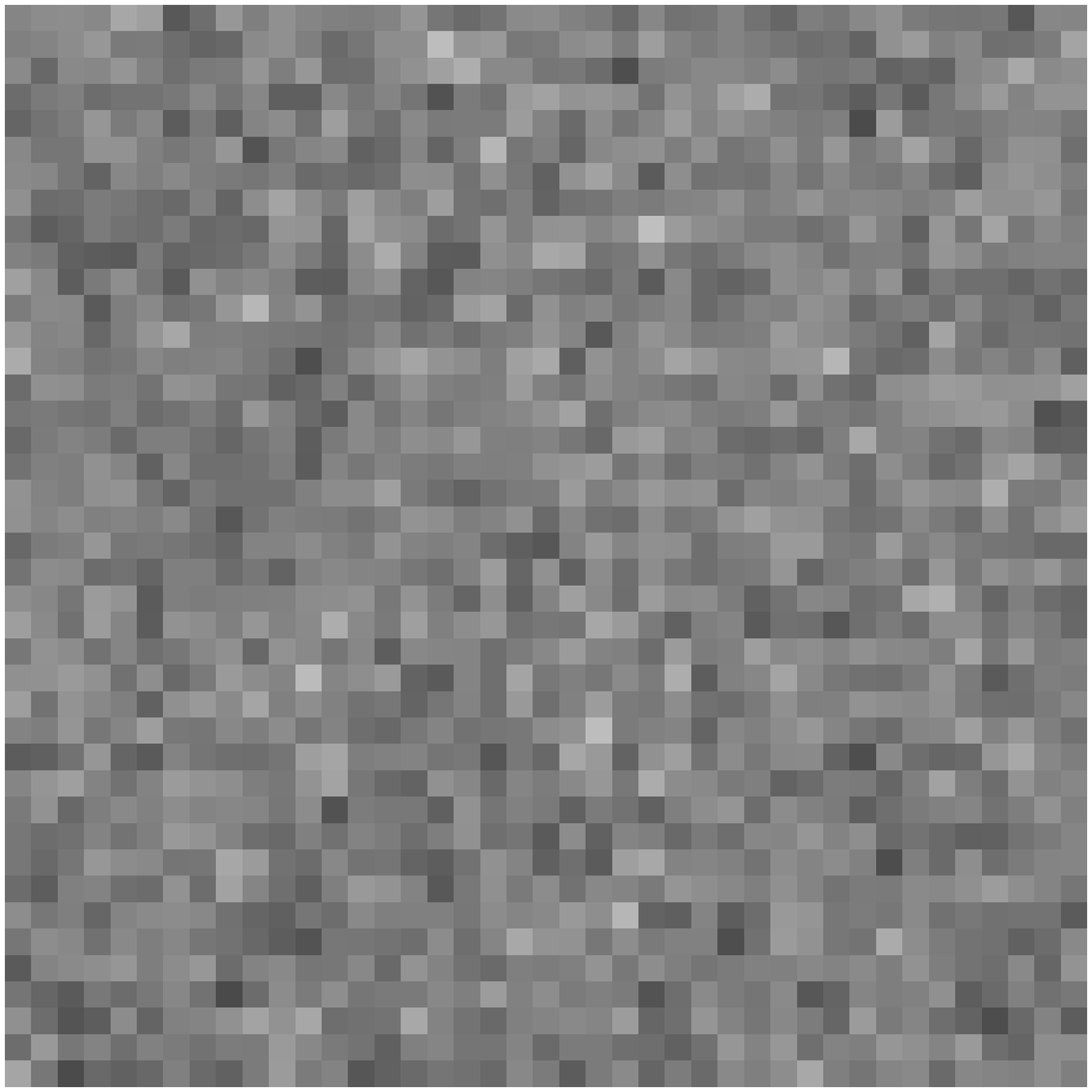}}}
  \centerline{\subfigure[3.6-$\mu$m image, $N=8$.]{
\includegraphics[width=0.22\textwidth]{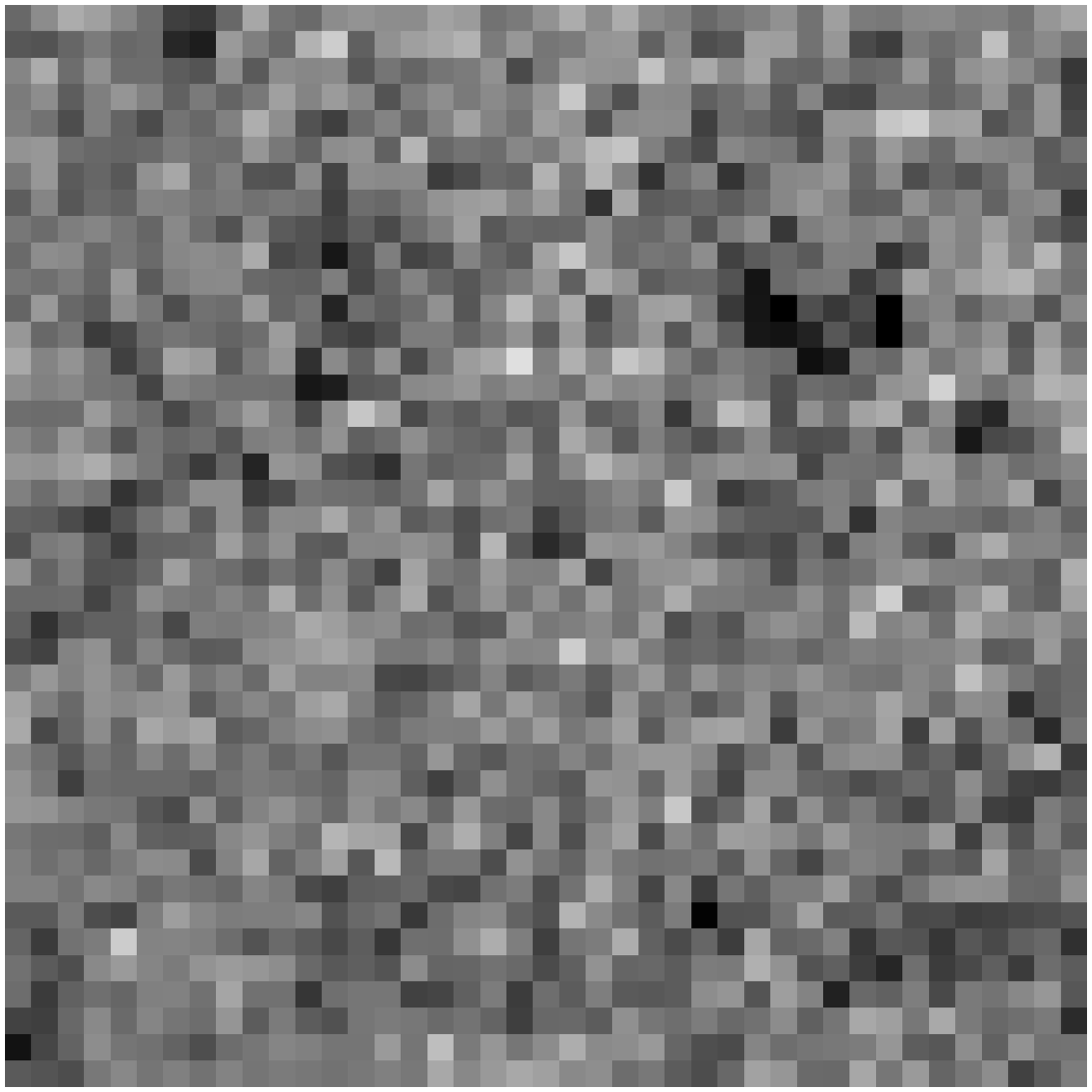}}
              \subfigure[4.5-$\mu$m image, $N=8$.]{
\includegraphics[width=0.22\textwidth]{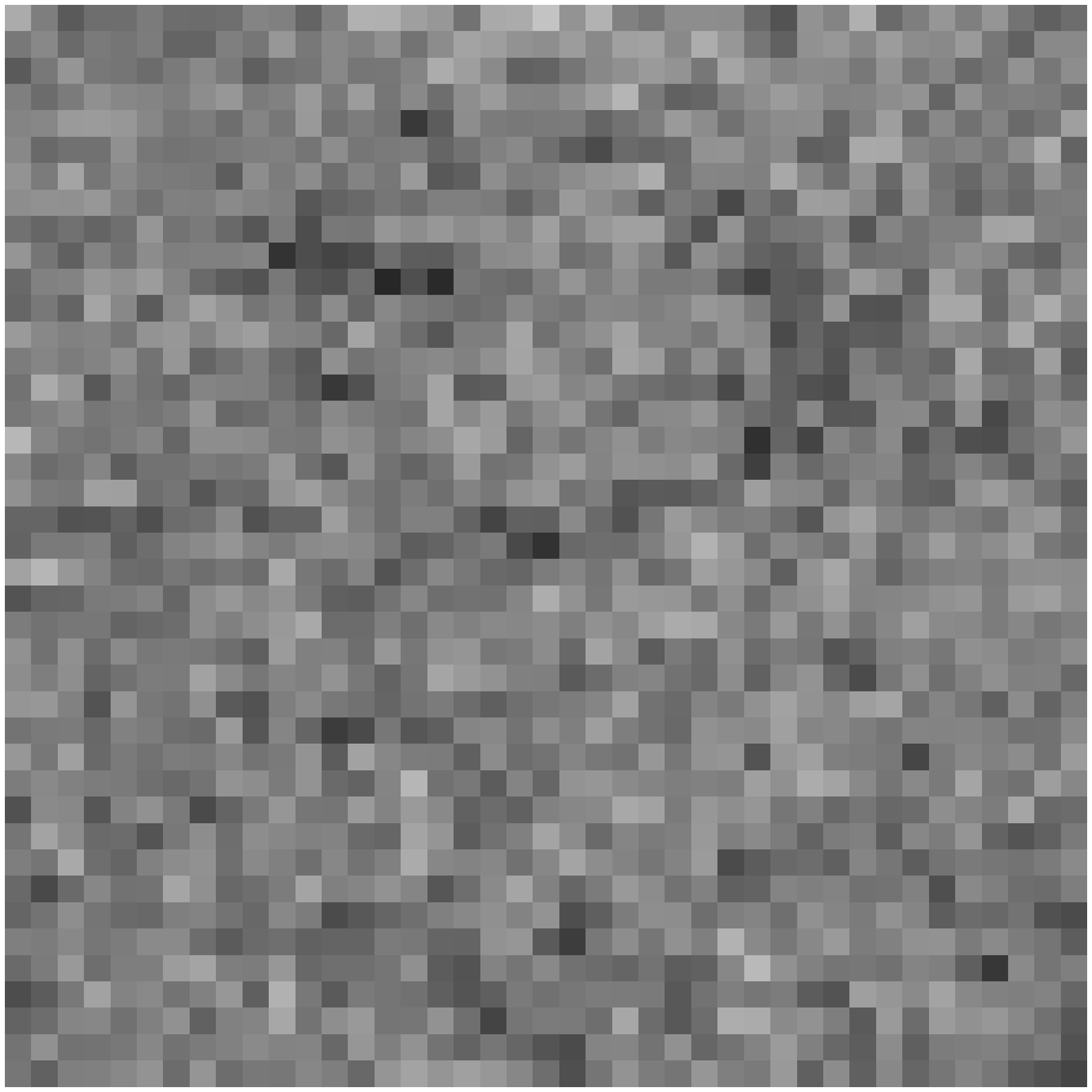}}
              \subfigure[5.8-$\mu$m image, $N=8$.]{
\includegraphics[width=0.22\textwidth]{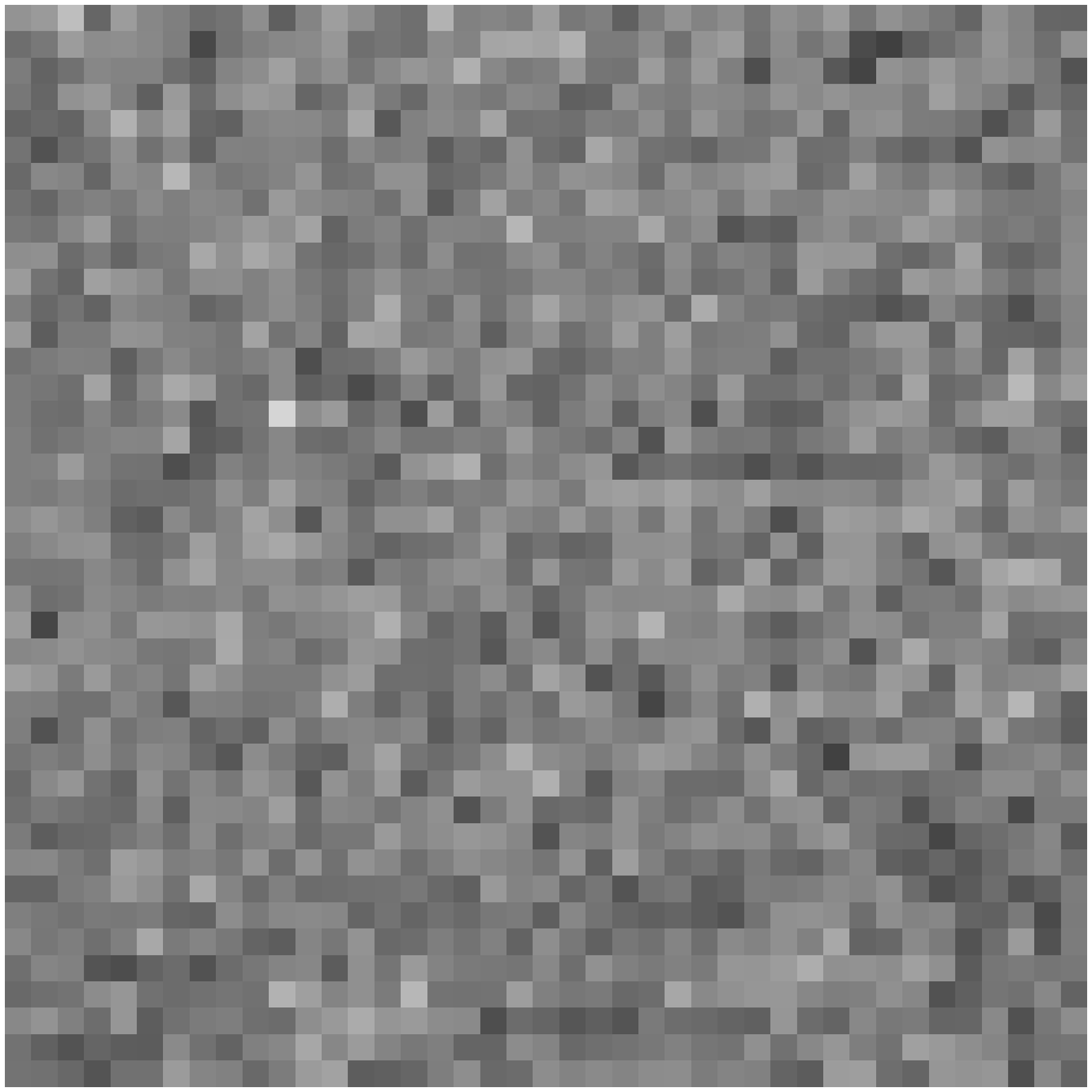}}
              \subfigure[8-$\mu$m image, $N=8$.]{
\includegraphics[width=0.22\textwidth]{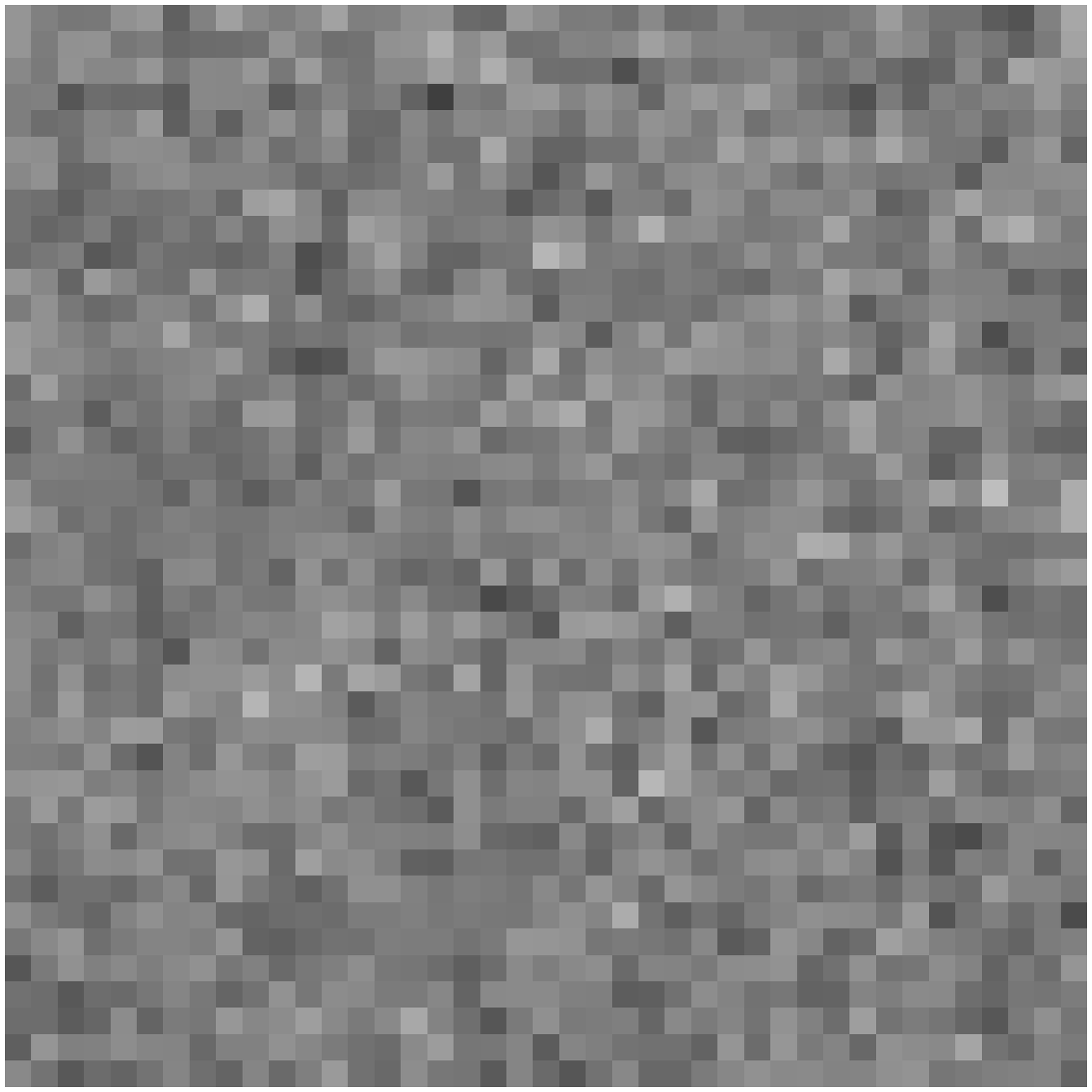}}}
    \caption{Stacked IRAC images.  The grey-scales for each image
    correspond to the range $\pm \sigma/\sqrt{N}$, for comparison with
    Fig.~\ref{fig:IFRSs}, where $\sigma$ = 2.9, 5, 20 and
    20~$\mu$Jy~pixel$^{-1}$ for the four bands.  The radio source
    locations are at the centre of each stacked image.  Images (a) to
    (d) show the stacked images in the four IRAC bands from all 14
    IFRSs, while (e) to (h) use only the 8 IFRS from
    Table~\ref{tab:IFRSproperties} which have not been classified as
    having a possible uncatalogued infrared counterpart.}
    \label{fig:stack}
\end{figure*}

Figs.~\ref{fig:stack}a to d show the stacked images for the four IRAC
bands, with the grey-scale varying between $\pm \sigma/\sqrt{14}$ for
the relevant IRAC noise level $\sigma$ (where $\sigma$ = 2.9, 5, 20,
20~$\mu$Jy for the four bands).  There is an increase in flux density
seen in the centre of the 3.6- and 4.5-$\mu$m images, while the 5.8-
and 8-$\mu$m images show no evidence for emission.  This is
confirmation that at least some of the potential uncatalogued infrared
counterparts are real.  Figs.~\ref{fig:stack}e to h show the stacked
images created by using only the eight IFRSs which were not identified
as having potential uncatalogued infrared counterparts -- the lower
number of sources used for the stacking makes these images look
noisier, but there is still a slight indication of an increase in
infrared flux density in the centre of the 3.6- and 4.5-$\mu$m images.

In order to account for the possibility of a detection near to the
noise level, we place $3\sigma/\sqrt{8}$ upper limits on the infrared
flux densities of these sources of 3.1~$\mu$Jy at 3.6~$\mu$m,
5.3~$\mu$Jy at 4.5~$\mu$m, and 21.2~$\mu$Jy at 5.8 and 8~$\mu$m, and
set a 24-$\mu$m limit of 210~$\mu$Jy (equal to the 24-$\mu$m catalogue
flux limit for the main survey region), and limits of 10 and 50~mJy
for the 70-$\mu$m and 160-$\mu$m data, again from the catalogue flux
density limits.

Star-forming galaxies are observed to follow a tight correlation
between their mid- or far-infrared flux density, $S_{\rm IR}$, and
their 1.4-GHz flux density, $S_{1.4}$, which can be quantified by the
logarithmic flux density ratio $q_{\rm IR} = {\rm log}_{10}(S_{\rm IR}
/ S_{1.4})$ \citep*[e.g.][hereafter Paper II]{Appleton04,Garn08IR}.
The non-detections of IFRSs in the 24 and 70-$\mu$m {\it Spitzer}
bands allows us to place upper limits of $q_{24} < -0.7$, and $q_{70}
< 1$ -- these values are significantly below the typical values seen
for star-forming galaxies of $q_{24} \sim 1$ and $q_{70} \sim 2$
\citep{Appleton04}, and are associated with AGN activity.

\subsection{Source counterparts at other wavelengths}
The NASA Extragalactic
Database\footnote{http://nedwww.ipac.caltech.edu/} (NED) was searched
to find existing information on the IFRSs.  The eight sources with the
greatest 1.4-GHz flux density were present in the 1.4-GHz FIRST
survey.  Four sources were present in the 1.4-GHz NRAO VLA Sky Survey
\citep[NVSS;][]{Condon98}, and four were present in the 1.4-GHz survey
of the xFLS region by \citet{Morganti04}.  We use the \citet{Condon03}
1.4-GHz data in preference to any of these surveys, due to the
combination of high sensitivity and resolution.  One source
(FLSGMRT~J171436.6$+$594456) has a radio counterpart within the
325-MHz Westerbork Northern Sky Survey \citep[WENSS;][]{Rengelink97}.
The flux density given in NED is $27.0\pm4.9$~mJy, which is in
agreement with the value of $30.8\pm0.4$~mJy obtained by extrapolating
the 610-MHz flux density and spectral index (see
section~\ref{sec:spectralindex}).

Eight of the sources had an optical counterpart within 3~arcsec in the
$R$-band images of \citet{Fadda04}.  The $R$-band magnitudes (Vega)
are listed in Table~\ref{tab:IFRSproperties} where appropriate, and
are close to the 50~per~cent completeness limit of the observations,
which was estimated to be $R = 24.5$.  Where no detection was found,
this completeness level was taken as the upper limit on $R$-band flux.
No other optical or infrared counterparts were found for any of the
IFRSs.

\subsection{Radio spectral index}
\label{sec:spectralindex}
One of the few tools available for determining the type of IFRS being
observed is their radio spectral index, due to the lack of detections
in other wavebands.  The spectral index of each source is listed in
Table~\ref{tab:IFRSproperties}, and appear to be uniformly distributed
between 0 and 1.4 -- a one-sample KS test gives a 99~per~cent chance
that the data are consistent with a uniform distribution, although the
low number of sources limits the significance of this conclusion.

We can use the spectral index as a discriminant to determine between
different potential source types which make up the IFRS population.
The lack of infrared detections, and VLBI detection of a compact radio
core in one source indicate that obscured star-forming galaxies are
unlikely to be the dominant source population -- this is reinforced by
the fact that only six of the sources have a spectral index in the
range $0.5<\alpha\leq 1$, close to the typical value for `normal'
star-forming galaxies of $\sim0.8$ \citep[e.g.][]{Condon92}.

Three of the IFRSs are Ultra Steep Spectrum (USS) sources, with
$\alpha > 1$.  Selecting sources with ultra-steep spectral indices is
one method for efficiently finding high-redshift radio galaxies
\citep[HzRGS,
  e.g.][]{Tielens79,Rottgering94,Blundell98,deBreuck00,Klamer06}.
There are a number of mechanisms proposed to explain the relationship
seen between spectral index and redshift (the $z$-$\alpha$
correlation) -- see \citet{Klamer06} for a recent discussion of these
-- with the high-redshift sources selected using this technique
typically being compact, luminous Fanaroff-Riley Type II
\citep[FRII;][]{Fanaroff74} radio galaxies.

The remaining five sources have flat spectra, with $\alpha \leq 0.5$.
A flat-spectrum source with compact morphology (see
section~\ref{sec:linearsize}) is an indication of synchrotron self
absorption occurring either due to core-dominated sources, or sources
with a number of overlapping optically-thick regions aligned close to
our line of sight \citep[e.g.][]{Jarvis02}.

\section{Modelling the IFRS population}
\label{sec:modelling}
\subsection{Linear size constraints}
\label{sec:linearsize}
\begin{figure}
  \begin{center}
    \includegraphics[width=0.45\textwidth]{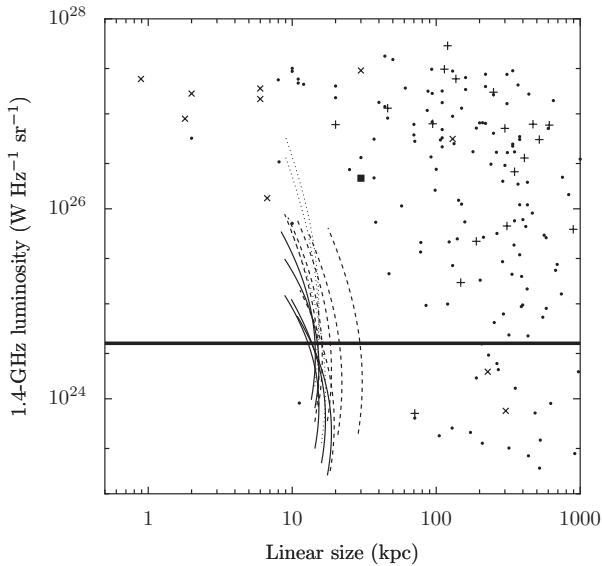}
    \caption{The luminosity -- linear size plot for sources in the
    3CRR catalogue, along with the tracks that each of the IFRSs would
    make across the plot if they were located at redshifts between
    $z=1$ (lower right) and $z=8$ (upper left).  Each line represents
    the track of one IFRS, and the linear sizes given by the IFRS
    tracks are upper limits.  Sources have been separated into
    flat-spectrum ($\alpha \leq0.5$; diagonal crosses for the 3CRR
    sources, solid lines for the IFRSs), steep-spectrum ($0.5< \alpha
    \leq 1$; dots for the 3CRR sources, dashed lines for the IFRSs)
    and USS sources ($\alpha >1$; upright crosses for the 3CRR
    sources, dotted lines for the IFRSs).  The flat-spectrum source
    3C273 is marked by the solid square.  The horizontal line marks
    the FRI/FRII divide at 1.4~GHz.}
    \label{fig:LinearSize}
  \end{center}
\end{figure}
We list the deconvolved angular diameters of each source in
Table~\ref{tab:IFRSproperties}, taken from the catalogue of
\citet{Condon03}.  From the larger diameter, we calculate upper limits
on the linear size of the IFRSs at a given redshift, using the
relationship between angular diameter-distance and $z$.  We model the
luminosity and maximum linear size for each source at redshifts
between 0 and 8, using the observed radio flux densities, and
$k$-correcting each source to obtain a rest-frame 1.4-GHz luminosity,
$L_{1.4}$, at each redshift, with a $k$-correction factor of
$(1+z)^{\alpha-1}$ (see e.g.\ Paper II).  Fig.~\ref{fig:LinearSize}
shows the tracks of maximum linear size against luminosity for the
IFRSs on a luminosity-size (P-D) diagram, for $1<z<8$, along with the
locations of sources in the Revised Revised Third Cambridge (3CRR)
catalogue of
\citet*{Laing83}\footnote{http://3crr.extragalactic.info/}, the
definitive list of bright radio sources in the northern sky at
178~MHz.  The 3CRR catalogue has been chosen for comparison to the
IFRS population due to the large amounts of photometry which are
available on a significant number of bright radio sources.  The break
luminosity between Fanaroff-Riley Class I and Class II sources
(FRI/FRII) is shown, adjusted to a 1.4-GHz value of
$3.84\times10^{24}$~W~Hz$^{-1}$~sr$^{-1}$ using $\alpha = 0.8$.  The
flat, steep and USS sources are shown separately.

The typical upper limit on linear size for the IFRSs is $\sim20$~kpc
at any redshift, and it is clear from Fig.~\ref{fig:LinearSize} that
the majority of 3CRR sources are too large to represent local versions
of the IFRSs.  In the following sections we model each class of source
in turn, using spectral energy distributions of suitable sources from
the 3CRR catalogue in order to demonstrate that IFRSs can be modelled
as being high-redshift versions of well-studied local sources.
Throughout this modelling, we fix the luminosity of the template 3CRR
sources to match the radio flux density constraints for each IFRS, and
use sources from Table~\ref{tab:IFRSproperties} which have not been
classified as having possible infrared counterparts where possible,
since they are the more extreme examples of IFRSs.

\subsection{Modelling the flat-spectrum sources}
\citet{Norris07} claimed that, based on their 1.4-GHz detection and
3.6-$\mu$m limit, the IFRS population may be represented by a
high-redshift version of 3C273.  This source is a radio-loud quasar
located at $z = 0.158339$ \citep{Strauss92}, with a single radio jet
visible with linear size $\sim30$~kpc \citep{Bahcall95}, 1.4-GHz flux
density of 45~Jy \citep*{Kellermann69}, and spectral index of
$\sim0.1$, giving it a luminosity of
$2.1\times10^{26}$~W~Hz$^{-1}$~sr$^{-1}$.  It is not actually in the
3CRR catalogue due to its low declination, but has similar properties
to the flat-spectrum IFRSs (although it is slightly larger than the
linear size constraints from Fig.~\ref{fig:LinearSize}).  We have a
  number of additional constraints on the spectra of IFRSs compared
  with \citet{Norris07}:
\begin{enumerate}
\item We have radio detections of J171453.8$+$594329 (the IFRS which
  fitted the spectra of 3C273 most successfully) at both 610~MHz and
  1.4~GHz.  The luminosity of 3C273 was reduced by a constant
  (redshift-dependent) factor in order to match these observations.
\item We required that the 1.4-GHz luminosity of the reduced version
  of 3C273 was great enough to remain above the FRI/FRII break
  luminosity.  If the luminosity was to be reduced below this level,
  the FRII SED of 3C273 would be modelling a radio galaxy with the
  luminosity of an FRI source, which would be inappropriate.
\item We place infrared upper limits on the IFRS population in the
  IRAC and MIPS bands, between observation-frame wavelengths of 3.6
  and 160~$\mu$m.
\item We have optical detections, or upper limits, in the $R$-band,
  which further constrains the redshift of each source.
\end{enumerate}

\begin{figure}
  \begin{center}
    \includegraphics[width=0.45\textwidth]{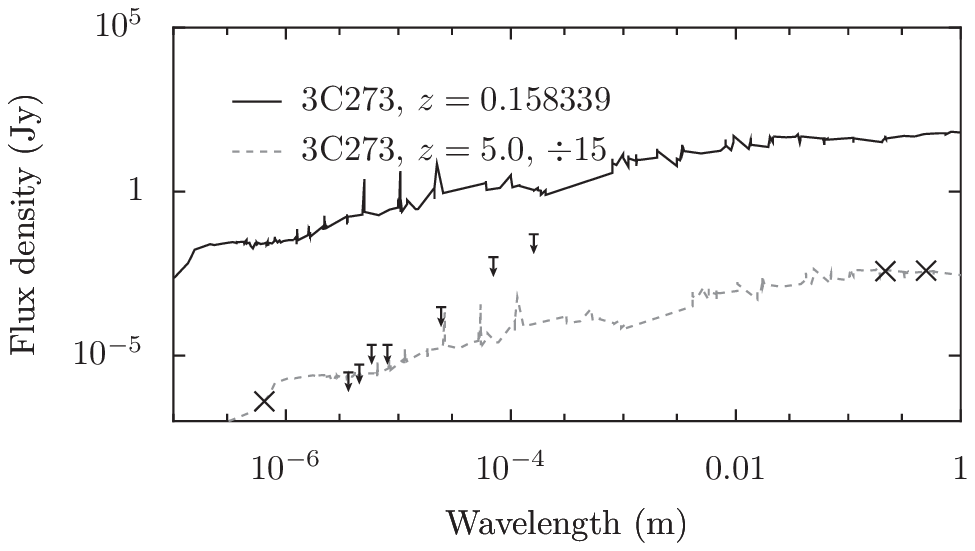}
    \caption{The spectral energy distribution of 3C273, as it is
    observed (black line), and as it would appear if 3C273 was placed
    at $z=5$ and reduced in luminosity by a factor of 15 (grey dashed
    line).  The detections and upper limits on flux density for the
    IFRS J171453.8$+$594329 are shown (diagonal crosses and upper
    limits) -- the SED of 3C273 would be below the infrared detection
    limits, and match the optical photometry when placed at $z=5$.}
    \label{fig:3c273}
  \end{center}
\end{figure}

We obtained photometry on 3C273 from NED for comparison with the IFRS
data, and placed it at redshifts between 0 and 8, to examine how the
SED that would be observed changes with redshift.
Fig.~\ref{fig:3c273} shows the observed SED of 3C273, and the SED that
would be observed if 3C273 was placed at $z=5$, and reduced in
luminosity by a factor of 15 (to a 1.4-GHz luminosity of
$1.4\times10^{25}$~W~Hz$^{-1}$~sr$^{-1}$).  This is a lower limit to
the redshift of the source (based upon fitting the optical/infrared
photometry).  Once the redshift is great enough that the SED shape can
be fitted successfully, there is a degeneracy between the redshift and
luminosity of the source, with the luminosity of 3C273 requiring a
smaller amount of scaling at greater redshifts.  The other
flat-spectrum IFRSs can also be modelled with the SED of 3C273, with
similar lower limits being placed on their redshift.  The compact
flat-spectrum sources in the 3CRR catalogue were also tested to see
whether they could successfully model the flat-spectrum IFRS
population, but less photometry was available on these sources, and
none of them was as good a fit to the data as 3C273.  Since the 3CRR
catalogue contains the brightest (low-frequency) sources in the sky,
it should not be suprising that the flat-spectrum sources contained
within it need to have their luminosity reduced significantly in order
to fit the IFRSs.  In carrying out the modelling in this section, we
have assumed that a reduction in luminosity by a factor of 15 does not
fundamentally alter the properties of the quasar being modelled -- it
would be preferable to use lower-luminosity sources for this
modelling, however there are few radio sources in the sky that have
been as well-studied at optical and infrared wavelengths as the 3CRR
catalogue.

\subsection{Modelling the steep-spectrum sources}
\label{sec:modelsteep}
There are six steep-spectrum sources with $0.5<\alpha\leq 1$ within
the IFRS sample, which follow very similar tracks in the P-D diagram.
The 3CRR catalogue contains several local sources with steep spectra
and linear size $<$20~kpc, with both FRI and FRII sources satisfying
the linear size requirements, and it is necessary to test both forms
of SED in order to determine which class of source best represents the
IFRS population.

\begin{figure}
  \begin{center}
    \includegraphics[width=0.45\textwidth]{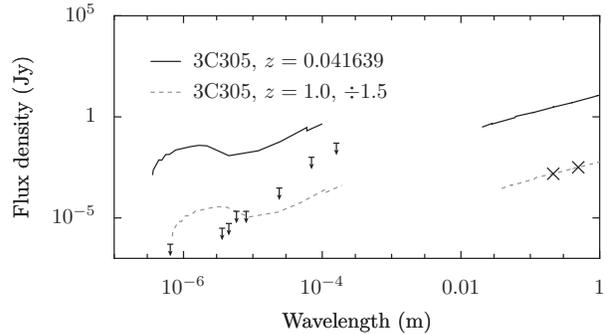}
    \caption{The spectral energy distribution of 3C305, as it
    is observed (black line), and as it would appear if 3C305 was
    placed at $z=1$ and reduced in luminosity by a factor of 1.5 (grey
    dashed line).  The detections and upper limits on flux density for
    the IFRS J172315.5$+$590919 are shown (diagonal crosses and upper
    limits) -- the FRI SED of 3C305 is unable to fit the infrared and
    optical data successfully.}
    \label{fig:3c305}
  \end{center}
\end{figure}

\begin{figure}
  \begin{center}
    \includegraphics[width=0.45\textwidth]{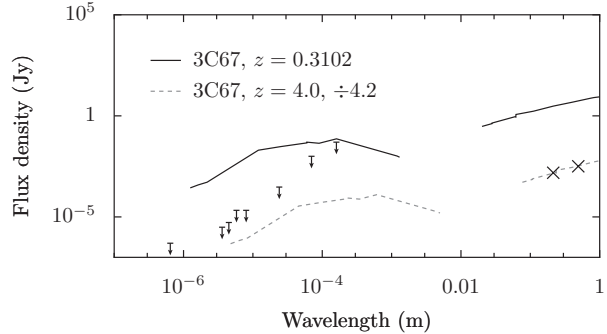}
    \caption{The spectral energy distribution of 3C67, as it is
    observed (black line), and as it would appear if 3C67 was placed
    at $z=4$ and reduced in luminosity by a factor of 4.2 (grey dashed
    line).  The detections and upper limits on flux density for the
    IFRS J172315.5$+$590919 are shown (diagonal crosses and upper
    limits) -- the FRII SED of 3C67 is able to fit the infrared and
    optical data successfully.}
    \label{fig:3c67}
  \end{center}
\end{figure}

Fig.~\ref{fig:3c305} shows the spectrum of 3C305, an FRI radio galaxy
at $z=0.041639$ \citep{Miller02}, which has been placed at $z=1$, and
reduced in luminosity slightly (by a factor of 1.5).  While the radio
detections and far-infrared flux limits can be matched, the
mid-infrared and optical limits placed on the IFRS J172315.5$+$590919
are inconsistent with the flux that would be observed from this
source.  No improvement was found to be possible at other redshifts.
In contrast to this, Fig.~\ref{fig:3c67} shows the spectrum of 3C67,
an FRII radio galaxy at $z=0.3102$ \citep{Hewitt91}, which has been
placed at $z=4$ and reduced in luminosity by a factor of 4.2.  All of
the available photometry can now be fitted successfully, demonstrating
that the steep-spectrum sources are better fitted by FRII templates
than by FRI templates.  A successful fit is also possible with 3C67
placed at lower redshifts (i.e.\ 3C67 at $z=3$ would require a
reduction in the total luminosity by a factor of $\sim9$ to fit the
radio data), but the origin of the optical and radio luminosities in
FRII galaxies is very different and this amount of luminosity scaling
may not be justified.  The lower limit on redshift for this source is
$z=2.5$, placed by requiring the 1.4-GHz radio luminosity to remain
above the FRI/FRII break.

\subsection{Modelling the ultra-steep spectrum sources}
\begin{figure}
  \begin{center}
    \includegraphics[width=0.45\textwidth]{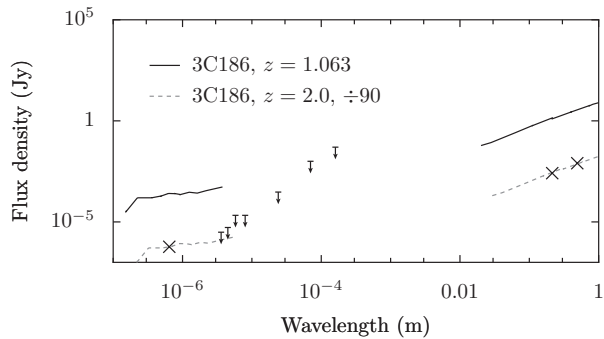}
    \caption{The spectral energy distribution of 3C186, as it
    is observed (black line), and as it would appear if 3C186 was
    placed at $z=2$ and reduced in luminosity by a factor of 90 (grey
    dashed line).  The detections and upper limits on flux density for
    the IFRS J171315.5$+$590302 are shown (diagonal crosses and
    limits) -- the SED of 3C186 is able to fit the radio and optical
    data, and the infrared limits at 3.6 and 4.5~$\mu$m (the only
    points where photometry on 3C186 is available).}
    \label{fig:3c186}
  \end{center}
\end{figure}

There are three sources within the IFRS sample with $\alpha > 1$.  One
of these, J171928.4$+$583927, has a spectral index of $1.04\pm0.09$
and is better fitted by the SED of 3C67 than to any of the 3CRR USS
sources, while the other two have spectral indices of 1.37 and 1.38.
The 3CRR catalogue only contains two compact sources with ultra-steep
spectra -- of the two, 3C186 \citep[an FRII source at
$z=1.063$;][]{Lynds66} has more available photometry, and
Fig.~\ref{fig:3c186} demonstrates that it is possible to fit the radio
and optical data for IFRS J171315.5$+$590302, along with the
3.6-$\mu$m IRAC infrared limit, by placing 3C186 at $z=2$ and
reducing its luminosity by a factor of 90.  As seen in previous
sections, placing the source at higher redshift allows this luminosity
scaling to be reduced while still being able to fit the observed
photometry -- at $z=4$ the luminosity of 3C186 needs to be reduced by
a factor of $\sim15$, while at $z=6$ the luminosity only needs to be
reduced by a factor of $\sim5$.
 
\subsection{Comparison with the `Optically Invisible' radio sources of
    \citet{Higdon05}} 

\citet{Higdon05} describe a population of sources which are
radio-bright but `optically invisible', with the majority of their
optically-invisible radio sources (OIRSs) also being undetected in the
mid-infrared with MIPS.  While the majority of their OIRSs are not
particularly radio-bright, there are two sources with 1.4-GHz flux
density $>1$~mJy, with no detections in optical images having $B_{W}
\sim 27.0$, $R \sim 25.7$, $I \sim 25.0$ (Vega magnitudes), or at
24~$\mu$m, with $S_{24} < 0.3$~mJy\footnote{There are two further
radio sources with $S_{1.4} > 1$~mJy and no optical detections,
outside the coverage area of MIPS, and one further source which is
described as an OIRS in \citet{Higdon05}, but no longer thought to be
one by \citet{Higdon08}.}.  \citet{Higdon05} conclude that these
sources are powered by AGN activity (principally based upon the
non-detections at 24~$\mu$m).  IRAC information is added to the OIRS
sample by \citet{Higdon08}, who find that at least 34~per~cent of the
sources do not have detections at 3.6~$\mu$m.  They conclude that the
OIRSs which are undetected by IRAC appear to represent a population of
powerful radio galaxies, at $z > 2$.

The Infrared-Faint and Optically-Invisible radio source populations
appear to be similar in nature, although with slightly different
selection effects.  While it remains to be confirmed that these
sources are high-redshift radio galaxies (potentially through
extremely deep multi-wavelength optical and infrared observations), it
is not necessary to require new, exotic source types in order to
explain the IFRS or OIRS population.

\section{Conclusions}
We confirm the existence of a class of bright radio sources which are
undetected in deep infrared images.  These sources are rare -- we find
a total of 14 sources with $S_{1.4} > 0.5$~mJy and no clear infrared
counterparts.  Ten of these have a flux density of greater than 1~mJy,
a source density of roughly 2.5~deg$^{-2}$, and comparable to the
source density of bright IFRSs found by \citet{Norris06} and
\citet{Middelberg08}.  There are eight sources with $S_{1.4}>0.5$~mJy
that show no evidence for possible infrared counterparts in the deep
IRAC images of the {\it Spitzer} extragalactic First Look Survey
field.

We use source stacking to place upper limits on the mid-infrared flux
densities of these sources, and use either detections or upper limits
in the $R$-band to further constrain the SED of each source.  By using
multi-frequency radio data, we demonstrate that the sources vary in
type between flat-spectrum quasars with $\alpha\sim0.1$, and USS radio
sources with $\alpha\sim1.4$, and should not be thought of as a single
source population.

The possibility that the Infrared-Faint Radio Source population is
made up of high-redshift luminous radio galaxies appears increasingly
likely -- the lack of detection in infrared images would seem to rule
out highly obscured star-forming galaxies, while the existence of
flat-spectrum sources implies that at least some of the sources have
an AGN origin.  We place upper limits on their overall linear size
through the knowledge of their maximum deconvolved angular size from
the \citet{Condon03} 1.4-GHz catalogue, and show that they must be
$<20$~kpc at any redshift, making them smaller than the majority of
bright radio sources in the local Universe, as given by the 3CRR
catalogue.  By looking at the locations of these sources in a P-D
diagram, for a range of redshifts, we identify bright local radio
sources with similar characteristics, and demonstrate that all three
classes of IFRS can be modelled as less-luminous versions of
well-studied Fanaroff-Riley Type II radio galaxies placed at high
redshift.

\section*{Acknowledgments}
TG thanks the UK STFC for a Studentship.  We thank the staff of the
GMRT who have made these observations possible.  The GMRT is operated
by the National Centre for Radio Astrophysics of the Tata Institute of
Fundamental Research, India.  This work has made use of the NASA/IPAC
Extragalactic Database (NED) which is operated by the Jet Propulsion
Laboratory, California Institute of Technology, under contract with
the National Aeronautics and Space Administration.

\bibliography{./References}

\begin{thebibliography}{}

\bibitem[\protect\citeauthoryear{Appleton et~al.,}{Appleton
  et~al.}{2004}]{Appleton04}
Appleton P.~N.,  et~al., 2004, ApJS, 154, 147

\bibitem[\protect\citeauthoryear{Bahcall, Kirhakos, Schneider, Davis, Muxlow,
  Garrington, Conway \& Unwin}{Bahcall et~al.}{1995}]{Bahcall95}
Bahcall J.~N.,  Kirhakos S.,  Schneider D.~P.,  Davis R.~J.,  Muxlow T. W.~B.,
  Garrington S.~T.,  Conway R.~G.,    Unwin S.~C.,  1995, ApJ, 452, L91

\bibitem[\protect\citeauthoryear{Becker, White \& Helfand}{Becker
  et~al.}{1995}]{Becker95}
Becker R.~H.,  White R.~L.,    Helfand D.~J.,  1995, ApJ, 450, 559

\bibitem[\protect\citeauthoryear{Bertin \& Arnouts}{Bertin \&
  Arnouts}{1996}]{Bertin96}
Bertin E.,  Arnouts S.,  1996, A\&AS, 117, 393

\bibitem[\protect\citeauthoryear{Beswick, Muxlow, Thrall, Richards \&
  Garrington}{Beswick et~al.}{2008}]{Beswick08}
Beswick R.~J.,  Muxlow T. W.~B.,  Thrall H.,  Richards A. M.~S.,    Garrington
  S.~T.,  2008, MNRAS, 385, 1143

\bibitem[\protect\citeauthoryear{Blundell, Rawlings, Eales, Taylor \&
  Bradley}{Blundell et~al.}{1998}]{Blundell98}
Blundell K.~M.,  Rawlings S.,  Eales S.~A.,  Taylor G.~B.,    Bradley A.~D.,
  1998, MNRAS, 295, 265

\bibitem[\protect\citeauthoryear{Condon}{Condon}{1992}]{Condon92}
Condon J.~J.,  1992, ARA\&A, 30, 575

\bibitem[\protect\citeauthoryear{Condon, Cotton, Yin, Perley, Taylor \&
  Broderick}{Condon et~al.}{1998}]{Condon98}
Condon J.~J.,  Cotton W.~D.,  Yin Q.~F.,  Perley R.~A.,  Taylor G.~B.,
  Broderick G.~B.,  1998, AJ, 115, 1693

\bibitem[\protect\citeauthoryear{Condon, Cotton, Yin, Shupe, Storrie-Lombardi,
  Helou, Soifer \& Werner}{Condon et~al.}{2003}]{Condon03}
Condon J.~J.,  Cotton W.~D.,  Yin Q.~F.,  Shupe D.~L.,  Storrie-Lombardi L.~J.,
   Helou G.,  Soifer B.~T.,    Werner M.~W.,  2003, AJ, 125, 2411

\bibitem[\protect\citeauthoryear{de Breuck, van Breugel, R{\" o}ttgering \&
  Miley}{de~Breuck et~al.}{2000}]{deBreuck00}
de Breuck C.,  van Breugel W.,  R{\" o}ttgering H. J.~A.,    Miley G.,  2000,
  A\&AS, 143, 303

\bibitem[\protect\citeauthoryear{Dunkley et~al.,}{Dunkley
  et~al.}{2008}]{Dunkley08}
Dunkley J.,  et~al., 2008, ApJS, submitted (astro-ph/0803.0586v1)

\bibitem[\protect\citeauthoryear{Fadda et~al.,}{Fadda  et~al.}{2006}]{Fadda06}
Fadda D.,  et~al., 2006, AJ, 131, 2859

\bibitem[\protect\citeauthoryear{Fadda, Jannuzi, Ford \&
  Storrie-Lombardi}{Fadda et~al.}{2004}]{Fadda04}
Fadda D.,  Jannuzi B.~T.,  Ford A.,    Storrie-Lombardi L.~J.,  2004, AJ, 128,
  1

\bibitem[\protect\citeauthoryear{Fanaroff \& Riley}{Fanaroff \&
  Riley}{1974}]{Fanaroff74}
Fanaroff B.~L.,  Riley J.~M.,  1974, MNRAS, 167, 31P

\bibitem[\protect\citeauthoryear{Fazio et~al.,}{Fazio  et~al.}{2004}]{Fazio04}
Fazio G.,  et~al., 2004, ApJS, 154, 10

\bibitem[\protect\citeauthoryear{Frayer et~al.,}{Frayer
  et~al.}{2006}]{Frayer06}
Frayer D.~T.,  et~al., 2006, AJ, 131, 250

\bibitem[\protect\citeauthoryear{Garn \& Alexander}{Garn \&
  Alexander}{2008}]{Garn08stacking}
Garn T.,  Alexander P.,  2008, MNRAS, submitted

\bibitem[\protect\citeauthoryear{Garn, Ford \& Alexander}{Garn
  et~al.}{2008}]{Garn08IR}
Garn T.,  Ford D.~C.,    Alexander P.,  2008, MNRAS, submitted

\bibitem[\protect\citeauthoryear{Garn, Green, Hales, Riley \& Alexander}{Garn
  et~al.}{2007}]{Garn07}
Garn T.,  Green D.~A.,  Hales S. E.~G.,  Riley J.~M.,    Alexander P.,  2007,
  MNRAS, 376, 1251

\bibitem[\protect\citeauthoryear{Hewitt \& Burbidge}{Hewitt \&
  Burbidge}{1991}]{Hewitt91}
Hewitt A.,  Burbidge G.,  1991, ApJS, 75, 297

\bibitem[\protect\citeauthoryear{Higdon et~al.,}{Higdon
  et~al.}{2005}]{Higdon05}
Higdon J.~L.,  et~al., 2005, ApJ, 626, 58

\bibitem[\protect\citeauthoryear{Higdon, Higdon, Willner, Brown, Stern,
  {Le~Floch'h} \& Eisenhardt}{Higdon et~al.}{2008}]{Higdon08}
Higdon J.~L.,  Higdon S. J.~U.,  Willner S.~P.,  Brown M. J.~I.,  Stern D.,
  {Le~Floch'h} E.,    Eisenhardt P.,  2008, ApJ, submitted
  (astro-ph/0806.2138v1)

\bibitem[\protect\citeauthoryear{Jarvis \& McLure}{Jarvis \&
  McLure}{2002}]{Jarvis02}
Jarvis M.~J.,  McLure R.~J.,  2002, MNRAS, 336, L38

\bibitem[\protect\citeauthoryear{Kellermann, Pauliny-Toth \&
  Williams}{Kellermann et~al.}{1969}]{Kellermann69}
Kellermann K.~I.,  Pauliny-Toth I. I.~K.,    Williams P. J.~S.,  1969, ApJ,
  157, 1

\bibitem[\protect\citeauthoryear{Klamer, Ekers, Bryant, Hunstead, Sadler \& de
  Breuck}{Klamer et~al.}{2006}]{Klamer06}
Klamer I.~J.,  Ekers R.~D.,  Bryant J.~J.,  Hunstead R.~W.,  Sadler E.~M.,
  de Breuck C.,  2006, MNRAS, 371, 852

\bibitem[\protect\citeauthoryear{Lacy et~al.,}{Lacy  et~al.}{2005}]{Lacy05}
Lacy M.,  et~al., 2005, ApJS, 161, 41

\bibitem[\protect\citeauthoryear{Laing, Riley \& Longair}{Laing
  et~al.}{1983}]{Laing83}
Laing R.~A.,  Riley J.~M.,    Longair M.~S.,  1983, MNRAS, 204, 151

\bibitem[\protect\citeauthoryear{Lonsdale et~al.,}{Lonsdale
  et~al.}{2003}]{Lonsdale03}
Lonsdale C.~J.,  et~al., 2003, PASP, 115, 897

\bibitem[\protect\citeauthoryear{Lynds, Hill, Heere \& Stockton}{Lynds
  et~al.}{1966}]{Lynds66}
Lynds C.~R.,  Hill S.~J.,  Heere K.,    Stockton A.~N.,  1966, ApJ, 144, 1244

\bibitem[\protect\citeauthoryear{Middelberg et~al.,}{Middelberg
  et~al.}{2008}]{Middelberg08}
Middelberg E.,  et~al., 2008, AJ, 135, 1276

\bibitem[\protect\citeauthoryear{Miller, Ledlow, Owen \& Hill}{Miller
  et~al.}{2002}]{Miller02}
Miller N.~A.,  Ledlow M.~J.,  Owen F.~N.,    Hill J.~M.,  2002, AJ, 123, 3018

\bibitem[\protect\citeauthoryear{Morganti, Garrett, Chapman, Baan, Helou \&
  Soifer}{Morganti et~al.}{2004}]{Morganti04}
Morganti R.,  Garrett M.~A.,  Chapman S.,  Baan W.,  Helou G.,    Soifer T.,
  2004, A\&A, 424, 371

\bibitem[\protect\citeauthoryear{Norris et~al.,}{Norris
  et~al.}{2006}]{Norris06}
Norris R.~P.,  et~al., 2006, AJ, 132, 2409

\bibitem[\protect\citeauthoryear{Norris, Tingay, Phillips, Middelberg, Deller
  \& Appleton}{Norris et~al.}{2007}]{Norris07}
Norris R.~P.,  Tingay S.,  Phillips C.,  Middelberg E.,  Deller A.,    Appleton
  P.~N.,  2007, MNRAS, 378, 1434

\bibitem[\protect\citeauthoryear{Rengelink, Tang, de Bruyn, Miley, Bremer, R{\"
  o}ttgering \& Bremer}{Rengelink et~al.}{1997}]{Rengelink97}
Rengelink R.~B.,  Tang Y.,  de Bruyn A.~G.,  Miley G.~K.,  Bremer M.~N.,  R{\"
  o}ttgering H. J.~A.,    Bremer M. A.~R.,  1997, A\&AS, 124, 259

\bibitem[\protect\citeauthoryear{Rieke et~al.,}{Rieke  et~al.}{2004}]{Rieke04}
Rieke G.~H.,  et~al., 2004, ApJS, 154, 25

\bibitem[\protect\citeauthoryear{R{\" o}ttgering, Lacy, Miley, Chambers \&
  Saunders}{R{\" o}ttgering et~al.}{1994}]{Rottgering94}
R{\" o}ttgering H. J.~A.,  Lacy M.,  Miley G.~K.,  Chambers K.~C.,    Saunders
  R.,  1994, A\&AS, 108, 79

\bibitem[\protect\citeauthoryear{Seymour, McHardy \& Gunn}{Seymour
  et~al.}{2004}]{Seymour04}
Seymour N.,  McHardy I.~M.,    Gunn K.~F.,  2004, MNRAS, 352, 131

\bibitem[\protect\citeauthoryear{Strauss, Huchra, Davis, Yahil, Fisher \&
  Tonry}{Strauss et~al.}{1992}]{Strauss92}
Strauss M.~A.,  Huchra J.~P.,  Davis M.,  Yahil A.,  Fisher K.,    Tonry J.,
  1992, ApJS, 83, 29

\bibitem[\protect\citeauthoryear{Tielens, Miley \& Willis}{Tielens
  et~al.}{1979}]{Tielens79}
Tielens A. G. G.~M.,  Miley G.~K.,    Willis A.~G.,  1979, A\&AS, 35, 153

\bibitem[\protect\citeauthoryear{Werner et~al.,}{Werner
  et~al.}{2004}]{Werner04}
Werner M.,  et~al., 2004, ApJS, 154, 1

\bibitem[\protect\citeauthoryear{White, Helfand, Becker, Glikman \& de
  Vries}{White et~al.}{2007}]{White07}
White R.~L.,  Helfand D.~J.,  Becker R.~H.,  Glikman E.,    de Vries W.,  2007,
  ApJ, 654, 99

\end{thebibliography}
\bibliographystyle{mn2e}
\label{lastpage}

\end{document}